\documentclass[iop]{emulateapj}
\usepackage{natbib}
\usepackage{graphicx}

\usepackage[dvipsnames]{color}
\usepackage{float}
\usepackage{longtable} 
\usepackage{textgreek}
\usepackage{amsmath}
\usepackage{fancyref}
\definecolor{orange}{RGB}{255,69,0}
\definecolor{green}{RGB}{0,128,0}
\definecolor{darkred}{RGB}{139,0,0}
\usepackage{adjustbox}
\usepackage{subfigure}
\usepackage{amssymb}
\usepackage{lineno}
\usepackage[colorlinks=true,
            linkcolor=green,
            urlcolor=magenta,
            citecolor=blue]{hyperref}
\usepackage{nameref}
\usepackage{array,multirow,makecell}
\usepackage{tabularx}

\begin{document}
\title
{Constraints on Proton Synchrotron Origin of  VHE Gamma Rays from the Extended Jet of AP Librae}
\author{Partha Pratim Basumallick, Nayantara Gupta}
\email{email:basuparth314@gmail.com}
\affil{Raman Research Institute,
    C.V.Raman Avenue, Sadashivanagar, Bangalore 560080, India\\
    }

\begin{abstract}
The multi-wavelength photon spectrum from the BL Lac object AP Librae extends from radio to TeV gamma rays.  The X-ray to very high energy gamma ray emission from the extended jet of this source has been modeled  earlier with inverse Compton (IC) scattering of relativistic electrons off the CMB photons. The IC/CMB model  requires the  kpc scale extended jet  to be highly collimated with bulk Lorentz factor close to 10. Here we discuss  the possibility of proton synchrotron origin of X-rays and gamma-rays from the extended jet with bulk Lorentz factor 3. This scenario requires extreme energy of protons $3.98\times10^{21}$ eV and high magnetic field 1 mG of the extended jet with jet power $\sim 5\times 10^{48}$ ergs/sec in particles and magnetic field (which is more than 100 times the Eddington's luminosity of AP Librae) to explain the very high energy gamma ray emission.  
Moreover, we have shown that X-ray emission from the extended jets of 3C 273 and PKS 0637-752 could be possible by proton synchrotron emission with jet powers  comparable to their Eddington's luminosities.
\end{abstract}


\section{Introduction}
\label{Section_1}  
AP Librae a low frequency peaked BL Lac, being at a redshift of  0.0486 (\cite{Disney et al.(1974)}), has been observed in radio to TeV gamma rays by several detectors. In comparison with other extragalactic sources which have been detected in X--rays(from their extended jets) [e.g., 3C 279, PKS 0637-752] AP Librae has the distinguishing feature of being also detected in VHE $\gamma$--rays. As a consequence of this uniqueness in the SED of AP Librae the approach to modelling it presents some very interesting opportunities of investigation especially in view of the fact that the VHE region of the spectra cannot be specifically attributed to a particular region inside the source (central core/ parsec scale jet/ extended jet).
\par 
The multi-wavelength spectra from BL Lacs are usually well described  by synchrotron self Compton (SSC) model. It is difficult to distinguish between core and jet emission from the multi-wavelength data. The high energy (HE) photons detected by \textit{Fermi}-LAT  in the 100 MeV-100GeV energy range and the very high energy photons (VHE) detected by H.E.S.S above 100 GeV (\cite{Abramowski et al.(2015)}) provide good statistics for detailed modeling of AP Librae.
 Although in 2013 the \textit{Fermi}-LAT data did show a flare with maximum flux 3.5 times above the quiescent state, no flare was indicated in the very high energy gamma ray data recorded by H.E.S.S.  from AP Librae at that time.  Due to poor angular resolution it was not possible to ascertain the location of gamma ray emission in this source. 
 \par The spectrum from AP Librae has been modeled by SSC and external Compton (EC) earlier (see \cite{Hervet et al.(2015)}; \cite{Sanchez et al.(2015)}).
The extended jet of AP Librae has been observed in radio and X-ray frequencies. This 14 kpc long and 4.8 kpc wide jet having similar morphologies in radio and X-rays (\cite{Kaufmann et al.(2013)}) has been modeled by inverse Compton emission by CMB photons (\cite{Sanchez et al.(2015)}; \cite{Zacharias and Wagner(2016)}).
 \par
 It is not possible to fit all the data from AP Librae with a simple one zone SSC model. More complicated scenarios  have been considered in previous works. 
Within a blob-in-jet SSC scenario \cite{Hervet et al.(2015)} have introduced many components of inverse Compton emission. They have added up the synchrotron and SSC photon fluxes from the blob and a parsec scale jet, EC photon flux from blob-jet and blob-broad line region interactions and the  second order SSC photon flux from the blob to get the observed flux. The jet power in thermal, non-thermal particles and magnetic field required in their model is comparable to the Eddington's luminosity  of AP Librae, which is $3.75\times 10^{46}$ ergs/sec. 
  \par
  A compact zone with Lorentz bulk factor 20 and an extended jet of 10kpc radius and Lorentz bulk factor 8 are the two emission regions of AP Librae in the study by \cite{Sanchez et al.(2015)}. The very high energy gamma rays are mostly produced by IC/CMB in the extended jet in their model.
  
\par
  In a more recent study \cite{Zacharias and Wagner(2016)} the multi-wavelength spectrum from AP Librae has been modeled with three zones, a blob, a parsec scale jet and a kpc scale extended jet.  The authors have considered the same value of the bulk Lorentz factor ($\Gamma=10$) for all these three zones. Also in this case, the jet power required is found to be comparable to the Eddington's luminosity.
\par
A lepto-hadronic model has been invoked for AP Librae (\cite{Petropoulou et al.(2016)}), where Bethe-Heitler interactions and photo-meson production inside the blob have been suggested as the origin of  GeV-TeV gamma rays. In this case  hour scale variability is expected in the very high energy gamma ray  data due to its compact production region. Moreover, the jet power required in this model  is ($10^{47}$--$10^{48}$ erg/sec), 10-100 times higher than the Eddington's luminosity.
The authors have considered two zones --- a compact blob and a parsec scale jet of Lorentz bulk factor  8. The larger the emission region the lower is the seed photon density, as a result  Bethe-Heitler pairs and photo-mesons are more unlikely to be produced inside the jet.

\begin{figure*}[htp]
\centering
\includegraphics[scale=1.2]{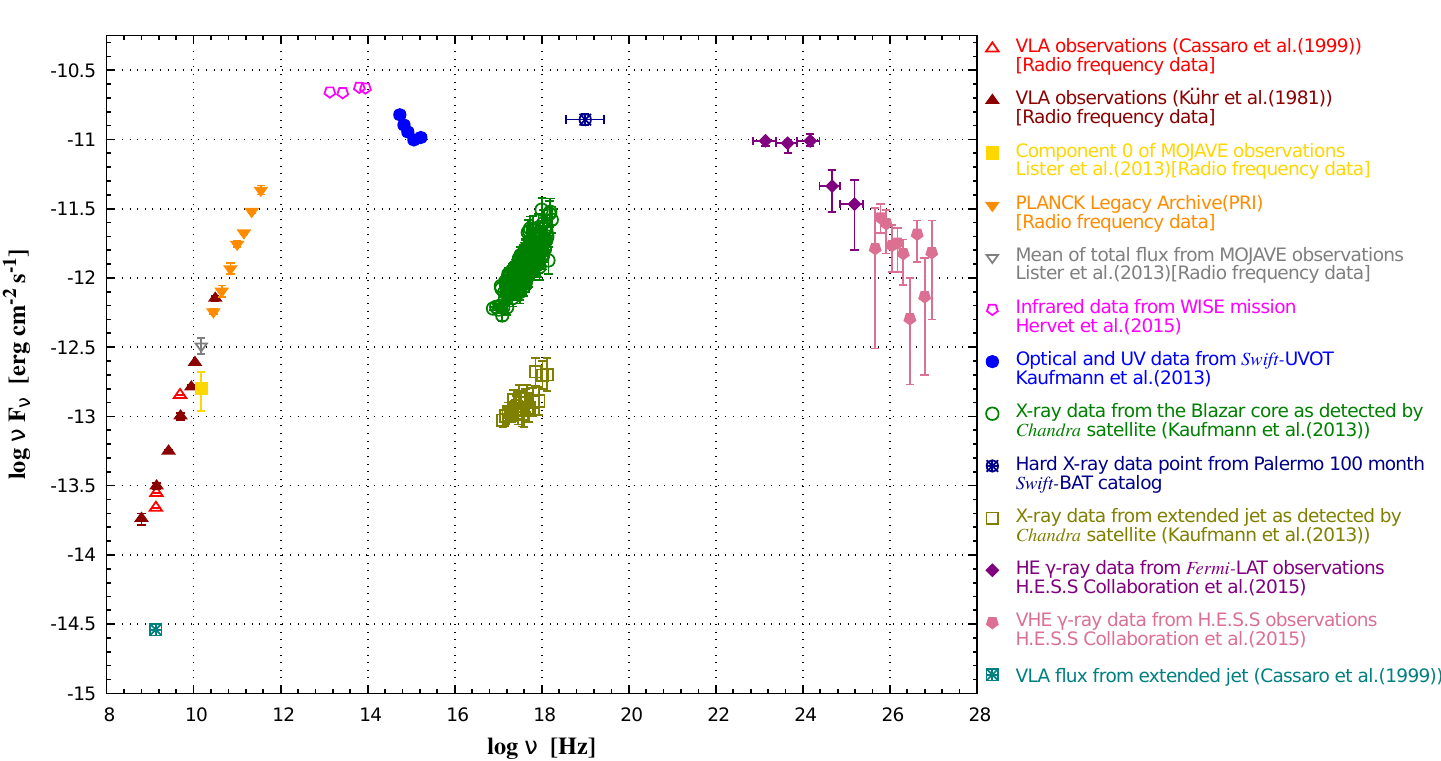}
\centering\caption{\textbf{
Quiescent state multiwavelength emission spectra of AP Librae}
}
\label{Figure-1}
\end{figure*} 

\par
Proton synchrotron emission from extended jets of Flat Spectrum Radio Quasars (FSRQs) has been studied earlier for Pictor A, 3C 120, 3C 273, PKS 0637-752 in \cite{Aharonian(2002)}, for 3C 273 in \cite{Kundu and Gupta(2014)} and for PKS 0637-752  in \cite{Bhattacharyya and Gupta(2016)}.
The authors  have compared the synchrotron loss and escape time scales of the ultrahigh energy protons with the age of the extended jet to determine the break energy in the proton spectrum and obtained the high energy photon flux from the synchrotron emission of these trapped protons.
\par
 In the present work the multi-wavelength data compiled in \cite{Zacharias and Wagner(2016)}  has been used to study the possibility of proton synchrotron origin of very high energy gamma ray  emission from the extended jet of AP Librae. We have considered a three zone model with a compact emission region or blob, a near parsec scale jet and an extended jet. 
 In the next section we discuss about the spectra of relativistic electrons, protons  in the compact region/blob near the core, in the near parsec scale jet, in the extended jet and their radiations to explain the spectral energy distributions (SEDs) of AP Librae.

 \section{Modeling the SED}
 \label{Section_2}
The multiwavelength data representing the quiescent state of AP Librae is presented in \autoref{Figure-1}. According to observations all the data points\footnote{Red open upward pointing triangle--\cite{Cassaro et al.(1999)};Maroon filled upward pointing triangle--\cite{Kuhr et al.(1981)};Yellow filled squares--\cite{Lister et al.(2013)};Orange downward pointing triangle--PLANCK Legacy Archive;Gray downward pointing triangle--\cite{Lister et al.(2013)}} in the radio frequency regime except the one at 1.36 GHz (\cite{Cassaro et al.(1999)}) are expected to originate from the central core region and parsec scale jet of the source. The data point at 1.36 GHz depicted by the boxed asterisk in teal (\autoref{Figure-1}) is attributed to the extended jet of AP Librae along with the X-ray data points depicted by the open squares in olive green. The green open circles on the other hand are the X-ray frequency data which have the blazar core region as their source of origin. Both the core and extended jet were detected in X-rays by the \textit{Chandra} satellite and were reported by \cite{Kaufmann et al.(2013)}. The hard X-ray data point depicted by navy blue circled asterisk was obtained from the Palermo 100 month \textit{Swift}-BAT catalog. The infrared frequency data shown as the pink coloured open pentagons are due to the WISE mission (\cite{Hervet et al.(2015)}) and the optical--UV frequency data depicted by the blue filled circles are from the \textit{Swift}-UVOT observations (\cite{Kaufmann et al.(2013)}). 

\par 

We consider three separate emission regions to explain the observed spectra of AP Librae--- \\i) a compact zone located near the blazar core which we refer to as Zone--1(Z1),  \\ii) a region of near parsec scale dimension referred to as Zone--2(Z2) and \\iii) an emission region in the extended jet of AP Librae referred to as Zone--3(Z3). \\The observed datapoints in the radio--optical--UV frequency region are easily explained by the electron synchrotron emission from Zone--1 \& Zone--2 and the radiation from the accretion disk of AP Librae. The X--ray frequency data depicted by the green open circles and olive green squares are explained by SSC emission from Zone--1 and proton synchrotron emission from Zone--3 respectively. However as the spatial resolution is comparatively poor for observations in the $>$ 100 MeV to TeV range it is not possible to distinguish either region as the source for the datapoints in that energy range. On the other hand as the VHE $\gamma$-rays detected by H.E.S.S (\cite{Abramowski et al.(2015)}) show no evidence of variability (as of now) it makes for a possible argument in favour of prescribing the extended jet of AP Librae as the emission region for the HE--VHE portion of the observed SED. Although according to the reports of \cite{Abramowski et al.(2015)} AP Librae indicates variability on the scale of a few days above the energy of 300 MeV it does not act as an essential factor in determining the emission region as the present paper deals with the quiescent state spectra of AP Librae. In any case as the spectra in the HE--VHE $\gamma$-ray region comprises of both proton synchrotron emission from the extended jet and SSC emission from Zone--2 (whose dimensions are consistent with a variability timescale of few days) our model should in principle account for the non quiescent state of AP Librae as well. The HE and VHE $\gamma$-ray data points are depicted by the purple diamonds and pale violet-red filled pentagons respectively. As it is expected that the data in the higher energy regime of the SED should suffer from absorption due to the extra galactic background radiation (EBL), the presented data points have been accordingly corrected for the intergalactic absorption using the EBL model of \cite{Franceschini et al.(2008)}.

Although \cite{Sanchez et al.(2015)} \& \cite{Zacharias and Wagner(2016)} have previously considered the extended jet as the source of HE and VHE emission of AP Librae, they have relied on inverse Compton emission mechanism to explain the higher energy emission in the observed SED. The present work on the other hand studies the possibility of proton synchrotron radiation from the extended jet of AP Librae to account for the HE--VHE $\gamma$-ray emission while simultaneously explaining the X-ray data originating from the extended jet. 

\subsection{Particle Spectra in Zone--1 $\&$ Zone--2}
   The electron spectra in Zone--1 and Zone--2 are broken power laws as they are cooling fast by synchrotron and self Compton emission.
\begin{equation}
\frac{dN_{el}(E_{el})}{dE_{el}} 
=A \left\{ \begin{array}{l@{\quad \quad}l}
{E_{el}^{-p_1}} &
E_{el}<E_{el}^{brk}\\{E_{el}^{brk}}
{E_{el}^{-p_1-1}} & E_{el}>E_{el}^{brk}
\end{array}\right.
\label{eqn_1}
\end{equation}

Out of the several input parameters required by the code, the extrema of the relativistic particles along with the so called break energy $E_{el}^{brk}$ are a prime factor in determining the generated spectrum. The break energy is determined by equating the cooling time $t_{cool}$ with $t_{esc}=\eta_{esc} \times \dfrac{R}{c}$, where $\dfrac{R}{c}$ is the light travel time in the emission region and $\eta_{esc}$ acts as a scaling factor. The cooling time scale of the electrons is given by the expression

 \begin{equation}
\dfrac{1}{t_{cool}}=\dfrac{1}{t_{synch}}+\dfrac{1}{t_{SSC}}
    \label{eqn_2}
    \end{equation}    which is a convolution of both the synchrotron cooling and SSC cooling timescales of the electrons. In a more detailed form \autoref{eqn_2} can be expressed as follows 

 \begin{equation}
 \dfrac{1}{t_{cool}}=\dfrac{4}{3} \sigma_{Th}^{el} \beta_{el}^{2} \gamma_{el} \dfrac{c}{m_{el} c^{2}} \left(U_{B}+U_{el}^{synch}\right)
    \label{eqn_3}
    \end{equation} 
where $\sigma_{Th}^{el}$ is the Thomson cross-section of electrons. The dimensionless speed of electrons $\beta_{el}\simeq 1$ and Lorentz factor  $\gamma_{el}=\dfrac{E_{el}}{m_{el}c^{2}}$. $U_{B}\left(=\dfrac{B^{2}}{8\pi}\right)$ and $U_{el}^{synch}$ are the energy densities\footnote{All the expressions \& values of energy densities are in the comoving frame of the emission regions throughout the paper unless otherwise mentioned.} of the magnetic field and the synchrotron radiation generated by the electrons respectively. \autoref{eqn_3} is valid only for scatterings in the Thomson regime. The Klein-Nishina effect does not play any significant role in our calculations. In our model the radius of the central core region is assumed to be $R_{Z1}=8.65\times 10^{16}$ cm while the radius of the parsec scale region is taken as $R_{Z2}=9.5\times 10^{17}$ cm. The magnetic field in the two zones are respectively $B_{Z1}=8.5$ mG and $B_{Z1}=1.72$ mG. The scaling factor $\eta_{esc}$ for Zone--1 and Zone--2 are respectively 62 and 63. As the value of $U_{el}^{synch}$ is sensitive to magnetic field, the energy range and also the energy density of the particle population, we have to adjust our parameter values before we can fix the break energy  of the electrons for a suitable choice of $\eta_{esc}$. 
 The relativistic protons are cooled by synchrotron emission inside the blob. 
The synchrotron emission from relativistic protons inside Zone--1 and Zone--2 is generated by assuming that the energy density ratio between protons and electrons is 1000:1. 
Proton cooling is not important in these zones and as the cooling time scale is long no break appears in the proton spectrum.
The maximum energy of the protons in the emission region is constrained by the condition: 
\begin{equation}
B\geqslant 30 \frac{E_p^{max} }{10^{19}} \frac{10^{15}}{R} \qquad \rm  {in\  Gauss}.
\label{eqn_4}
\end{equation}
which ensures that for our choice of \textit{B} the Larmor radii of protons having energy $E_p^{max}$ (in eV) does not exceed the chosen value of \textit{R}. It is important to mention in this context that although AP Librae exhibits intra-day scale (sometimes even as small as 20 min {\cite{Miller et al.(1974)}}) (\cite{Carini et al.(1991)};\cite{Webb et al. (1988)}) variability in the optical frequency range, we do not consider it as a constraining factor of \textit{R} owing to the fact that the data being modeled represents a flux averaged state of the source. However according to the approximate relation   
\begin{equation}
R{\leq}\frac{c\Delta t_{obs}\delta}{1+z}
\label{eqn_5}
\end{equation}
a variability timescale ($\Delta t_{obs}$) of 1 day for the Doppler factor $\delta = \frac{1}{\Gamma (1-\beta \cos \theta)}=12.3$ gives the radius of the emission region as $\sim 3\times 10^{16}$ cm. Similarly for the \textit{Fermi}-LAT observed 6 day variability observed in frequencies above 300 MeV the radius of the emission region should be $\sim 2 \times 10^{17}$ cm. However taking into account the fact that \autoref{eqn_5} might lead to large errors while estimating the dimensions of the emitting region (\cite{Protheroe(2002)}) our estimate of $R_{Z1}=8.65\times 10^{16}$ cm and $R_{Z2}=9.5\times 10^{17}$ cm should be admissible even when modelling the optical frequency data with day/intra-day scale variability and the HE $\gamma$-ray frequency data with week scale variability. 

\subsection{Particle Spectra in Extended Jet}
The electron  spectrum in the extended jet  (Zone--3)  also follows a broken power law given by \autoref{eqn_1}. 
The electron population in the extended jet radiates primarily via synchrotron emission (unlike the blob where SSC cooling is also a major contributor). The break energy of electrons $E_{brk,jet}^{el}$ is calculated by equating the synchrotron cooling time ($t_{synch}^{el}$) with the extended jet lifetime ($t_{jet}$). 

\begin{equation}
t_{synch}^{el}=t_{jet} 
\Rightarrow \dfrac{\gamma_{el} m_{el} c^2}{\dfrac{4}{3} \sigma_{Th}^{el} c U_{B} \gamma_{el}^2 \beta_{el}^2}=t_{jet}
\label{eqn_6}
\end{equation} 
The age of the extended jet $t_{jet}$ is assumed to be of the order of $10^5-10^7$ yrs. We note that the justification for assuming this age is to allow the ultrahigh energy protons to cool down by synchrotron emission during the lifetime of the jet.
   \par    
   The proton synchrotron emission from the extended jet of AP Librae is considered as the origin of the HE--VHE $\gamma$-rays in our model. 
    A broken power law proton spectrum is required to fit the $\gamma$- ray data.
   \begin{equation}
\frac{dN_{p}(E_{p})}{dE_{p}} 
=A \left\{ \begin{array}{l@{\quad \quad}l}
{E_{p}^{-p_1}} &
E_{p}<E_{p}^{brk}\\{{E_{p}^{brk}}^{(p_2-p_1)}}
{E_{p}^{-p_2}} & E_{p}>E_{p}^{brk}
\end{array}\right.
\label{eqn_7}
\end{equation}
  
 We consider energy dependent diffusive escape  of particles from the kpc scale extended jet and not the much smaller Doppler boosted compact regions  because the particles escape from the compact regions at a much faster rate, implying no significant amount of diffusion.
  \par
   We compare the synchrotron cooling time with the age of the extended jet and the escape timescale of the protons to fix the break energy.  The resulting proton spectrum is characterized by a change in the spectral index from $p_1$ before the break to $p_2$ after the break, in Bohm diffusion limit $p_2=p_1+1$.
   Also as the particles, both electrons and protons, are accelerated in the same region whether it be the blob or the extended jet, the spectral indices for both particle populations ($p_1$) before cooling, have been kept the same for the specific emission regions.
   The maximum energy of the protons on the other hand are constrained by the condition in \autoref{eqn_4}. The minimum energy of the protons is a free parameter and set to values of the order $\sim 10^{16}-10^{17}$ eV. Although the minimum energy values are a consequence of the modelling requirements of the observed SED,  higher values of the minimum energy  reduce the  jet power appreciably. 
 \par  
   
   The escape timescale and the synchrotron cooling time of the protons are given by \autoref{eqn_8} \& \autoref{eqn_9} respectively\footnote{Refer to \cite{Aharonian(2002)} for a detailed discussion}.
   \begin{equation}
t_{esc,Bohm}\simeq 4.2\times 10^5 \eta^{-1} B_{\rm mG} R^{2}_{\rm kpc} (E_{p}/10^{19}\rm eV)^{-1} \quad{\rm yrs}
\label{eqn_8}
\end{equation}
\begin{equation}
t_{synch}\simeq 1.4\times 10^{7} B^{-2}_{\rm{mG}} (E_{p}/10^{19} \rm eV)^{-1} \quad\rm{yrs}
\label{eqn_9}
\end{equation} 
In \autoref{eqn_8}, $\eta$ is the gyrofactor which assumes the value of 1 in the Bohm diffusion limit. $B_{\rm mG}$ is the ambient magnetic field (expressed in mG units) in the extended jet and $R_{\rm kpc}=\dfrac{R}{3.08\times 10^{21}}$ where \textit{R} is the radius of the emission zone (expressed in cm units) in the extended jet. When modelling the data in the Bohm diffusion regime we assume the value of $t_{jet}=2\times 10^7$ yrs. 
In our model in Bohm diffusion regime $E_{p}^{brk}=2.51\times 10^{18}$ eV for which the values of $t_{esc}$ \& $t_{synch}$ are $2.53\times 10^{7}$ yrs \& $5.6\times 10^{7}$ yrs respectively, which are comparable to the extended jet lifetime of $2\times10^7$ yrs.  
 
\par
We also consider the Kolmogorov and Kraichnan diffusion regimes to study the proton synchrotron model, the diffusion time scale is given in \autoref{eqn_10} 
\begin{equation}
t_{esc}\simeq \dfrac{R}{c} \left(\dfrac{E_p}{E_{free}}\right)^{-\alpha}
\label{eqn_10}
\end{equation}
where $E_{free}=E^*B_4R_{14}$ with $E^*=3\times 10^{20}$ eV, $B_4=\dfrac{B}{10^4}$ Gauss and $R_{14}=\dfrac{R}{10^{14}}$ cm respectively and $\alpha = \frac{1}{3}$ \&  $\frac{3}{5}$ for Kolmogorov and Kraichnan diffusion timescales.
 Spectral indices in Kolmogorov model are $p_{1, Kol}$ \& $p_{2, Kol} = p_{1, Kol}+\frac{1}{3}$ and in Kraichnan model $p_{1, Kra}$ \& $p_{2, Kra} = p_{1, Kra}+\frac{3}{5}$. In the Kolmogorov diffusion regime the escape timescale is $2.65\times 10^5$ yrs whereas that in the Kraichnan diffusion regime is $2.1\times 10^6$ yrs. The break energy for the electron population is calculated  assuming a jet lifetime of $2\times 10^5$ yrs and $2\times10^6$ yrs for the Kolmogorov and Kraichnan diffusion regimes respectively. The various values of the parameters used in the three diffusion regimes are presented in \autoref{Table-1}. 
 \par
  Although considering the different diffusive escape timescales does not affect the energetics and the parameter values are also similar nevertheless it does serve the purpose of showing that, inspite of large variations in the escape timescales (which are compared with the jet lifetime to ascertain the break energy) the values of the other parameters are hardly affected. Moreover as there are widely different estimates of the jet lifetime ($\sim 10^7-10^8$ yrs (\cite{Aharonian(2002)}) \& $\sim 10^5$ yrs (\cite{Kusunose and Takahara(2016)}) the assumption of different diffusive escape timescales allows us to consider comparable estimates of the jet lifetime. The fact that similar parameter values are required for each case is only found as a result of our study.
 \par
  If we assume energy independent escape of protons from the extended jet then the escape time is $t_{esc}=\eta_{esc}\times \left(\dfrac{R}{c}\right)$. The break energy in the proton spectrum is determined in this case by equating the synchrotron time scale to the escape time scale. We find that this scenario requires a very high value of the scaling factor $\eta_{esc}=4409$, compared to the compact regions. It is worth noticing in this context that assuming energy independent escape timescale doesn't require changing our parameter values for proton synchrotron spectrum from extended jet.
 \begin{figure}[H]
\centering
\includegraphics[width=.5\textwidth]{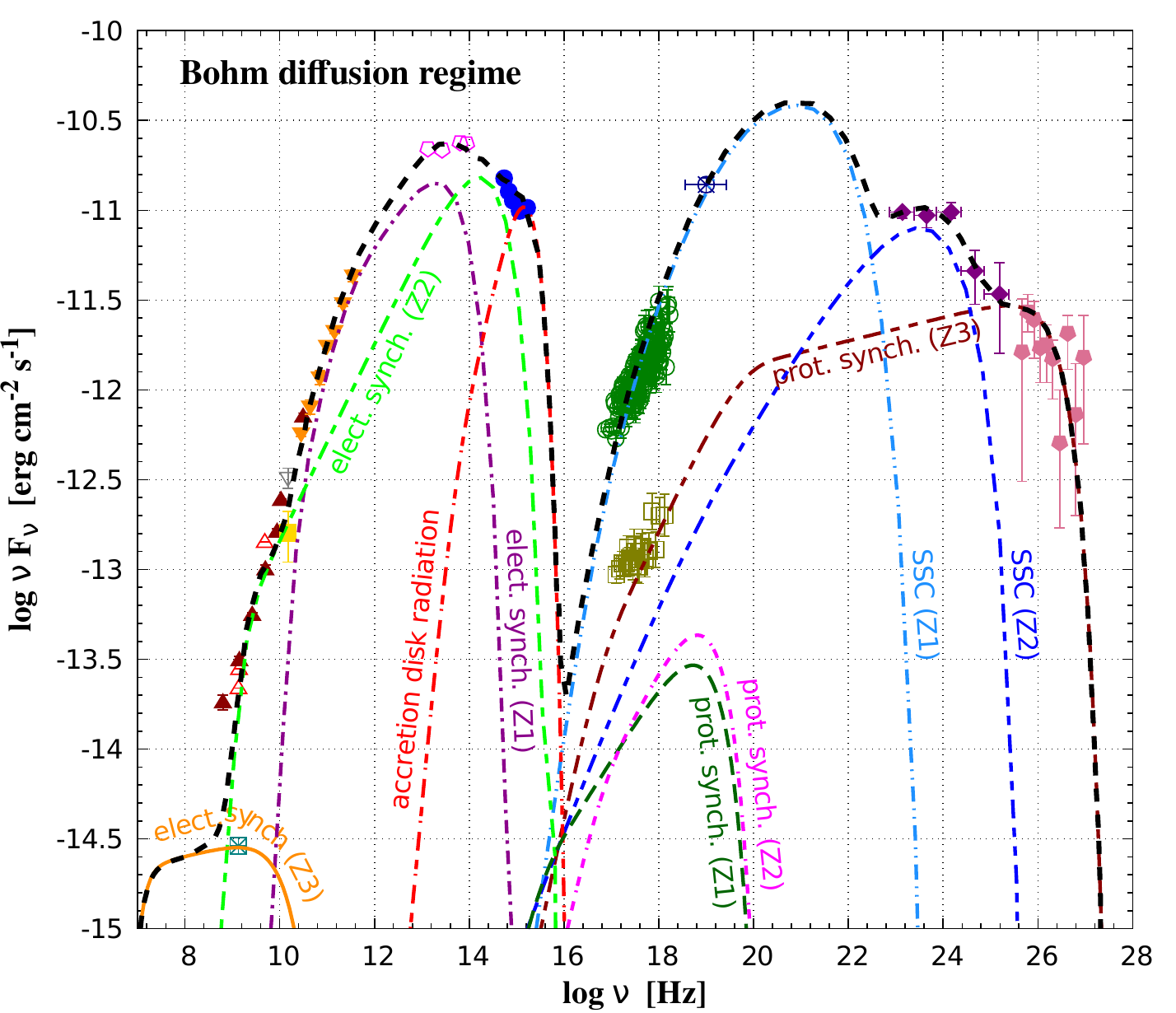}
\caption{\footnotesize{Multiple zone modeling of the observed steady state spectra of AP Librae.\\\\ \textit{Linetypes} -- Purple single dot long dashed line depicts electron synchrotron emission from the compact blob designated as zone--1(Z1). Lime green single short dashed long dashed line depicts the electron synchrotron emission from the parsec scale jet desinated as zone--2(Z2). Orange continuous line depicts the electron synchrotron emission from the extended jet designated as zone--3(Z3). Single dot short dashed long dashed line depicts the blackbody radiation from the accretion disk. Cyan double dotted long dashed line depicts the SSC emission from zone--1(Z1). Dark green long dashed line depicts the proton synchrotron emission from zone--1(Z1). Blue double short dashed long dashed line depicts the SSC emission from zone--2(Z2). Magenta single dot double short dashed line depicts the proton synchrotron emission from zone--2(Z2). Dark red double short dashed double long dashed line depicts the proton synchrotron emission from zone--3(Z3). Black single short dashed line depicts the overall emission spectra of the quiescent state multiwavelength data of AP Librae. The code developed by \cite{Krawczynski et al.(2004)} is used to generate the spectra from different emission mechanisms.}}
\label{Figure-2}
\end{figure}   
 \begin{figure}[H]
\centering
\includegraphics[width=.5\textwidth]{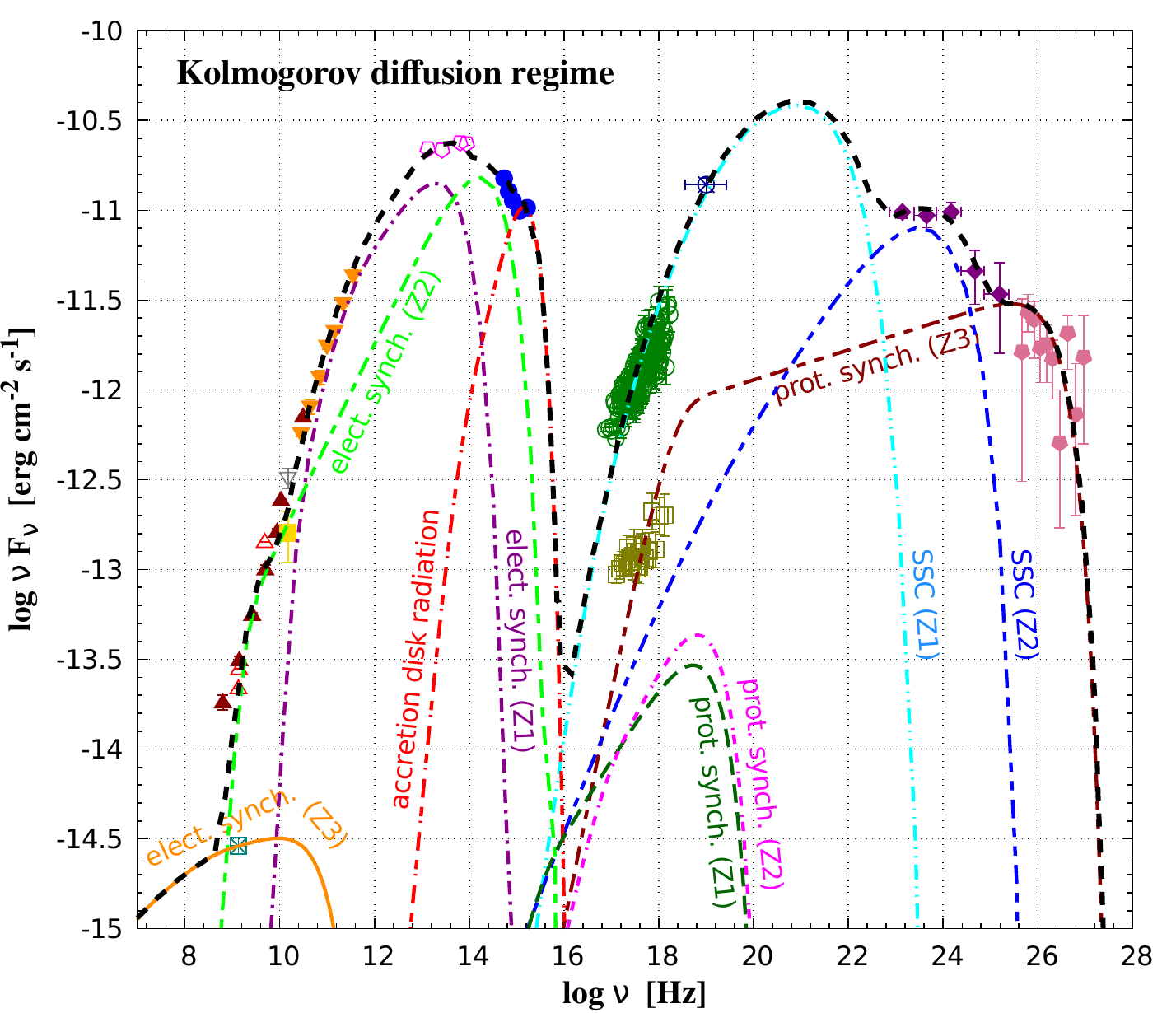}
\caption{\footnotesize{Same as in \autoref{Figure-2}.}}
\label{Figure-3}
\end{figure} 

 \begin{figure}[H]
\centering
\includegraphics[width=.5\textwidth]{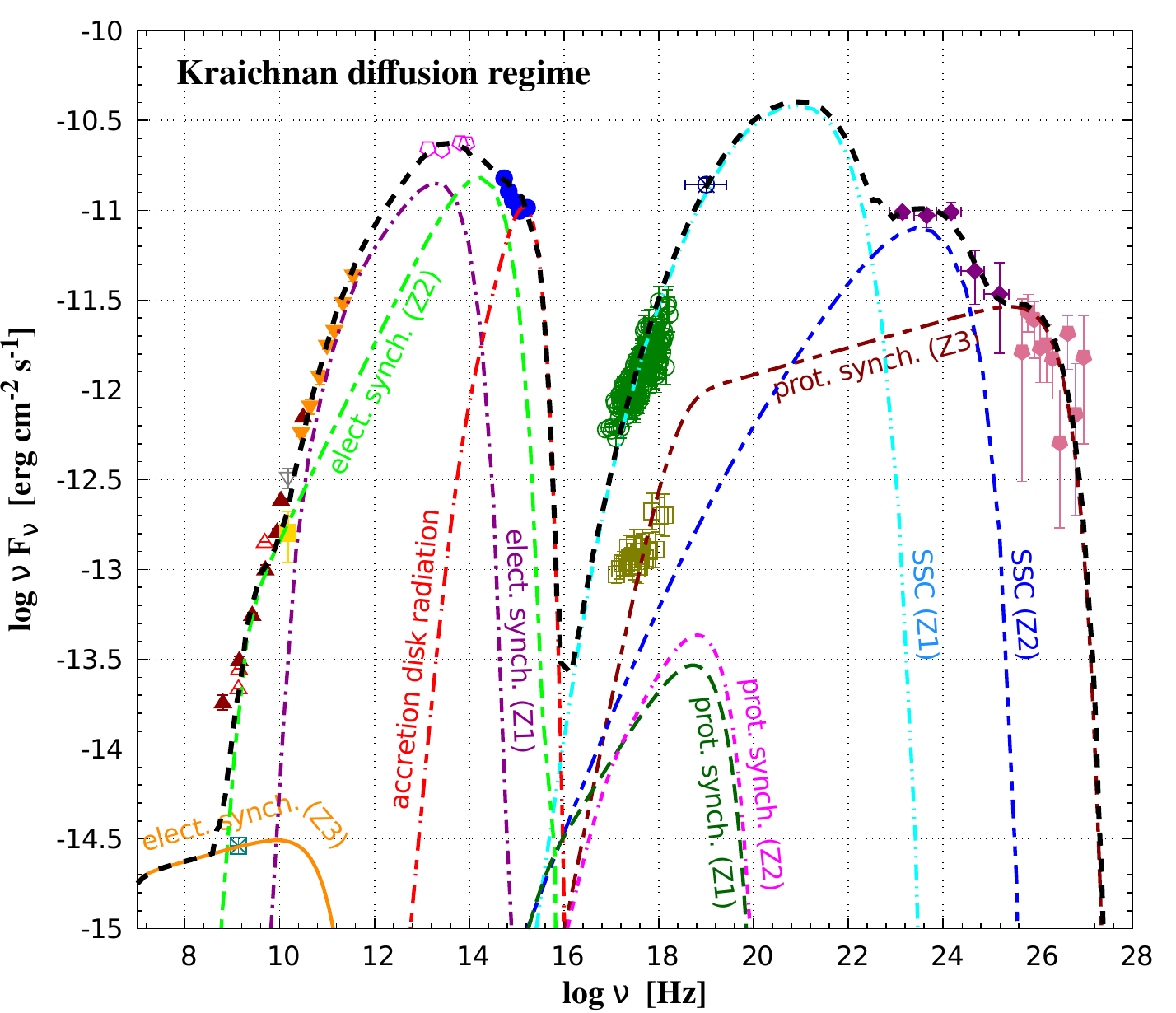}
\caption{\footnotesize{Same as in \autoref{Figure-2}.}}
\label{Figure-4}
\end{figure}

 \begin{table*}
\setlength\extrarowheight{5pt}

\centering
\caption{}{\textbf{Parameters for multiple zone modeling of the multiwavelength data from the quiescent state emission of AP Librae. The different diffusion regimes are assumed for the emission from extended jet.}}
\begin{center}
\scalebox{1.0}{
\begin{adjustbox}{center}
\begin{tabular}{|c|c|c|c|c|c|c|c|c|c|c|c|c|}

\hline

\textbf{Fitted} & \multirow{2}{*}{\textbf{Region}} & \multirow{2}{*}{\textbf{Emission\footnote{SSC emission from Zone-3 is found to be negligible.}\footnote{The proton synchrotron emission spectrum from Zone--1 $\&$ Zone--2 follow the simple power law whereas the proton synchrotron emission from Zone--3 follows a broken power law spectrum with the spectral indices being defined by the three different diffusion regimes as mentioned earlier.}}} & \multirow{2}{*}{\boldmath{$\Gamma$}} & \multirow{2}{*}{\boldmath{$\theta_{obs}$}} & \multirow{2}{*}{\boldmath{$\delta$}} & \multirow{2}{*}{\boldmath{$B$} {\bf mG}} & \multirow{2}{*}{\boldmath{$R$} {\bf cm}} & \multirow{2}{*}{\boldmath{$E_{min}$} {\bf eV}} & \multirow{2}{*}{\boldmath{$E_{max}$} {\bf eV}} &\multirow{2}{*}{\boldmath{$E_{break}$} {\bf eV}} & \multirow{2}{*}{\boldmath{$p_{1}$}} & \multirow{2}{*}{\boldmath{$p_{2}$}} \\
\textbf{SED}& & & & & & & & & & & & \\

\hline

\parbox[t]{2.5mm}{\multirow{6}{*}{\rotatebox[origin=c]{90}{\autoref{Figure-2}}}} \parbox[t]{2.5mm}{\multirow{6}{*}{\rotatebox[origin=c]{90}{Bohm diffusion}}}\parbox[t]{2.5mm}{\multirow{6}{*}{\rotatebox[origin=c]{90}{regime}}}& \multirow{2}{*}{Zone-1} &$e^-$ {synch}+SSC & \multirow{2}{*}{7}  & \multirow{2}{*}{3$^{\circ}$} & \multirow{2}{*}{12.3}  & \multirow{2}{*}{8.5} & \multirow{2}{*}{8.65$\times 10^{16}$} & 3.55$\times 10^{8}$  & 6.31$\times 10^{9}$  & 5.64$\times 10^{9}$  & 2.15  & 3.15 
   
\\\cline{3-3}\cline{9-13}

& & $p^+$ synch & & & & & & 3.16$\times 10^{15}$ & 2.45$\times 10^{17}$ & -- & 2.15 & --  

\\\cline{2-13}

& \multirow{2}{*}{Zone-2} & $e^-$ synch+SSC & \multirow{2}{*}{7}  & \multirow{2}{*}{3$^{\circ}$} & \multirow{2}{*}{12.3}  & \multirow{2}{*}{1.72} & \multirow{2}{*}{9.5$\times 10^{17}$} & 2.24$\times10^{7}$  & 3.55$\times 10^{10}$  & 3.51$\times 10^{10}$  & 1.9  & 2.9    
 
\\\cline{3-3}\cline{9-13}

& & $p^+$ synch & & & & & & 3.16$\times 10^{16}$ & 5.45$\times 10^{17}$ & -- & 1.9 & -- 

\\\cline{2-13}

& \multirow{2}{*}{Zone-3} & $e^-$ synch & \multirow{2}{*}{3}  & \multirow{2}{*}{5.5$^{\circ}$} & \multirow{2}{*}{5.42}  & \multirow{2}{*}{1} & \multirow{2}{*}{1.2$\times 10^{22}$} & 6.31$\times10^{5}$  & 3.98$\times 10^{8}$  & 6.5$\times 10^{5}$  & 1.87  & 2.87    
 
\\\cline{3-3}\cline{9-13}

& & $p^+$ synch & & & & & & 6.31$\times 10^{16}$ & 3.98$\times 10^{21}$ & 2.51$\times 10^{18}$ & 1.87 & 2.87  

\\\cline{2-13}

\hline

\parbox[t]{2.5mm}{\multirow{6}{*}{\rotatebox[origin=c]{90}{\autoref{Figure-3}}}} \parbox[t]{2.5mm}{\multirow{6}{*}{\rotatebox[origin=c]{90}{Kolmogorov diffusion}}}\parbox[t]{2.5mm}{\multirow{6}{*}{\rotatebox[origin=c]{90}{regime}}}& \multirow{2}{*}{Zone-1} &$e^-$ {synch}+SSC & \multirow{2}{*}{7}  & \multirow{2}{*}{3$^{\circ}$} & \multirow{2}{*}{12.3}  & \multirow{2}{*}{8.5} & \multirow{2}{*}{8.65$\times 10^{16}$} & 3.55$\times 10^{8}$  & 6.31$\times 10^{9}$  & 5.64$\times 10^{9}$  & 2.15  & 3.15 
   
\\\cline{3-3}\cline{9-13}

& & $p^+$ synch & & & & & & 3.16$\times 10^{15}$ & 2.45$\times 10^{17}$ & -- & 2.15 & --  

\\\cline{2-13}

& \multirow{2}{*}{Zone-2} & $e^-$ synch+SSC & \multirow{2}{*}{7}  & \multirow{2}{*}{3$^{\circ}$} & \multirow{2}{*}{12.3}  & \multirow{2}{*}{1.72} & \multirow{2}{*}{9.5$\times 10^{17}$} & 2.24$\times10^{7}$  & 3.55$\times 10^{10}$  & 3.51$\times 10^{10}$  & 1.9  & 2.9    
 
\\\cline{3-3}\cline{9-13}

& & $p^+$ synch & & & & & & 3.16$\times 10^{16}$ & 5.45$\times 10^{17}$ & -- & 1.9 & -- 

\\\cline{2-13}

& \multirow{2}{*}{Zone-3} & $e^-$ synch & \multirow{2}{*}{3}  & \multirow{2}{*}{5.5$^{\circ}$} & \multirow{2}{*}{5.42}  & \multirow{2}{*}{1} & \multirow{2}{*}{1.2$\times 10^{22}$} & 1.78$\times10^{6}$  & $10^{9}$  & 6.31$\times 10^{7}$  & 2.5  & 2.83    
 
\\\cline{3-3}\cline{9-13}

& & $p^+$ synch & & & & & & 3.8$\times 10^{17}$ & 3.98$\times 10^{21}$ & 3.98$\times 10^{17}$ & 2.5 & 2.83 
 
\\\cline{2-13}

\hline

\parbox[t]{2.5mm}{\multirow{6}{*}{\rotatebox[origin=c]{90}{\autoref{Figure-4}}}} \parbox[t]{2.5mm}{\multirow{6}{*}{\rotatebox[origin=c]{90}{Kraichnan diffusion}}}\parbox[t]{2.5mm}{\multirow{6}{*}{\rotatebox[origin=c]{90}{regime}}}& \multirow{2}{*}{Zone-1} &$e^-$ {synch}+SSC & \multirow{2}{*}{7}  & \multirow{2}{*}{3$^{\circ}$} & \multirow{2}{*}{12.3}  & \multirow{2}{*}{8.5} & \multirow{2}{*}{8.65$\times 10^{16}$} & 3.55$\times 10^{8}$  & 6.31$\times 10^{9}$  & 5.64$\times 10^{9}$  & 2.15  & 3.15 
   
\\\cline{3-3}\cline{9-13}

& & $p^+$ synch & & & & & & 3.16$\times 10^{15}$ & 2.45$\times 10^{17}$ & -- & 2.15 & --  

\\\cline{2-13}

& \multirow{2}{*}{Zone-2} & $e^-$ synch+SSC & \multirow{2}{*}{7}  & \multirow{2}{*}{3$^{\circ}$} & \multirow{2}{*}{12.3}  & \multirow{2}{*}{1.72} & \multirow{2}{*}{9.5$\times 10^{17}$} & 2.24$\times10^{7}$  & 3.55$\times 10^{10}$  & 3.51$\times 10^{10}$  & 1.9  & 2.9    
 
\\\cline{3-3}\cline{9-13}

& & $p^+$ synch & & & & & & 3.16$\times 10^{16}$ & 5.45$\times 10^{17}$ & -- & 1.9 & -- 

\\\cline{2-13}

& \multirow{2}{*}{Zone-3} & $e^-$ synch & \multirow{2}{*}{3}  & \multirow{2}{*}{5.5$^{\circ}$} & \multirow{2}{*}{5.42}  & \multirow{2}{*}{1} & \multirow{2}{*}{1.2$\times 10^{22}$} & 1.41$\times10^{6}$  & $10^{9}$  & 6.17$\times 10^{6}$  & 2.25 & 2.85    
 
\\\cline{3-3}\cline{9-13}

& & $p^+$ synch & & & & & & 3.8$\times 10^{17}$ & 3.98$\times 10^{21}$ & 7.08$\times 10^{17}$ & 2.25 & 2.85  

\\\cline{2-13}

\hline

\end{tabular}
\end{adjustbox}}
\end{center}
\label{Table-1}
\end{table*}

 \begin{table*}
\setlength\extrarowheight{5pt}

\centering
\caption{}{\textbf{Particle and magnetic energy densities in individual zones along with their contribution to the jet power for multiple zone modeling of the multiwavelength data of quiescent state emission from AP Librae.}}
\begin{center}
\scalebox{1.0}{
\begin{tabular}{|c|c|c|c|c|c|c|c|c|c|}

\hline

\textbf{Fitted} & \multirow{2}{*}{{\bf Region}}& \multirow{2}{*}{{\bf Particle}} & \boldmath{$u_{part.}^{'} $} & \boldmath{$u_{B}^{'}\left(=B^2/8\pi\right)$}  & \multirow{2}{*}{\boldmath{$R$} {\bf cm}} & \multirow{2}{*}{\boldmath{$\Gamma$}}    & \boldmath{$P_{part.,Z}^{'}$}   & \boldmath{$P_{B,Z}^{'}$}   & \boldmath{$P_{total,Z}^{'}$}     \\

\textbf{SED} &  &  & \boldmath{$\rm ergs/cm^3 $} & \boldmath{$\rm ergs/cm^3 $}  & & & {\bf ergs/sec}& {\bf ergs/sec} & {\bf ergs/sec}       \\

\hline

\parbox[t]{2.5mm}{\multirow{6}{*}{\rotatebox[origin=c]{90}{\autoref{Figure-2}}}} \parbox[t]{2.5mm}{\multirow{6}{*}{\rotatebox[origin=c]{90}{Bohm diffusion}}}\parbox[t]{2.5mm}{\multirow{6}{*}{\rotatebox[origin=c]{90}{regime}}} & \multirow{2}{*}{Zone-1} & Electron & $1.65\times 10^{-2}$  & \multirow{2}{*}{$2.87\times 10^{-6}$} & \multirow{2}{*}{$8.65\times10^{16}$}  & \multirow{2}{*}{7} & $5.7\times 10^{44}$ & \multirow{2}{*}{$9.92\times10^{40}$} & \multirow{2}{*}{$5.7\times10^{47}$}

\\\cline{3-4}\cline{8-8}

 &  & Proton & $1.65\times 10^{1}$  &  &   &  &$5.7\times 10^{47}$  &  &   
   
 \\\cline{2-10}
 
 & \multirow{2}{*}{Zone-2} & Electron & $9\times 10^{-5}$  & \multirow{2}{*}{$1.18\times 10^{-7}$} & \multirow{2}{*}{$9.5\times10^{17}$}  & \multirow{2}{*}{7} & $3.75\times 10^{44}$ & \multirow{2}{*}{$4.9\times10^{41}$} & \multirow{2}{*}{$3.75\times10^{47}$}

\\\cline{3-4}\cline{8-8}

 &  & Proton & $9\times 10^{-2}$  &  &   &  &$3.75\times 10^{47}$  &  &     
 
 \\\cline{2-10}
 
 & \multirow{2}{*}{Zone-3} & Electron & $1.9\times 10^{-15}$  & \multirow{2}{*}{$3.98\times 10^{-8}$} & \multirow{2}{*}{$1.2\times10^{22}$}  & \multirow{2}{*}{3} & $2.32\times 10^{41}$ & \multirow{2}{*}{$4.85\times10^{48}$} & \multirow{2}{*}{$4.86\times10^{48}$}

\\\cline{3-4}\cline{8-8}

 &  & Proton & $2.1\times 10^{-11}$  &  &   &  &$2.56\times 10^{45}$  &  &   \\ 
  
 \hline

\parbox[t]{2.5mm}{\multirow{6}{*}{\rotatebox[origin=c]{90}{\autoref{Figure-3}}}} \parbox[t]{2.5mm}{\multirow{6}{*}{\rotatebox[origin=c]{90}{Kolmogorov diffusion}}}\parbox[t]{2.5mm}{\multirow{6}{*}{\rotatebox[origin=c]{90}{regime}}} & \multirow{2}{*}{Zone-1} & Electron & $1.65\times 10^{-2}$  & \multirow{2}{*}{$2.87\times 10^{-6}$} & \multirow{2}{*}{$8.65\times10^{16}$}  & \multirow{2}{*}{7} & $5.7\times 10^{44}$ & \multirow{2}{*}{$9.92\times10^{40}$} & \multirow{2}{*}{$5.7\times10^{47}$}

\\\cline{3-4}\cline{8-8}

 &  & Proton & $1.65\times 10^{1}$  &  &   &  &$5.7\times 10^{47}$  &  &   
   
 \\\cline{2-10}
 
 & \multirow{2}{*}{Zone-2} & Electron & $9\times 10^{-5}$  & \multirow{2}{*}{$1.18\times 10^{-7}$} & \multirow{2}{*}{$9.5\times10^{17}$}  & \multirow{2}{*}{7} & $3.75\times 10^{44}$ & \multirow{2}{*}{$4.9\times10^{41}$} & \multirow{2}{*}{$3.75\times10^{47}$}

\\\cline{3-4}\cline{8-8}

 &  & Proton & $9\times 10^{-2}$  &  &   &  &$3.75\times 10^{47}$  &  &     
 
 \\\cline{2-10}
 
 & \multirow{2}{*}{Zone-3} & Electron & $3.1\times 10^{-16}$  & \multirow{2}{*}{$3.98\times 10^{-8}$} & \multirow{2}{*}{$1.2\times10^{22}$}  & \multirow{2}{*}{3} & $3.78\times 10^{40}$ & \multirow{2}{*}{$4.85\times10^{48}$} & \multirow{2}{*}{$4.86\times10^{48}$}

\\\cline{3-4}\cline{8-8}

 &  & Proton & $2.1\times 10^{-11}$  &  &   &  &$2.56\times 10^{45}$  &  &   \\

 \hline

\parbox[t]{2.5mm}{\multirow{6}{*}{\rotatebox[origin=c]{90}{\autoref{Figure-4}}}} \parbox[t]{2.5mm}{\multirow{6}{*}{\rotatebox[origin=c]{90}{Kraichnan diffusion}}}\parbox[t]{2.5mm}{\multirow{6}{*}{\rotatebox[origin=c]{90}{regime}}} & \multirow{2}{*}{Zone-1} & Electron & $1.65\times 10^{-2}$  & \multirow{2}{*}{$2.87\times 10^{-6}$} & \multirow{2}{*}{$8.65\times10^{16}$}  & \multirow{2}{*}{7} & $5.7\times 10^{44}$ & \multirow{2}{*}{$9.92\times10^{40}$} & \multirow{2}{*}{$5.7\times10^{47}$}

\\\cline{3-4}\cline{8-8}

 &  & Proton & $1.65\times 10^{1}$  &  &   &  &$5.7\times 10^{47}$  &  &   
   
 \\\cline{2-10}
 
 & \multirow{2}{*}{Zone-2} & Electron & $9\times 10^{-5}$  & \multirow{2}{*}{$1.18\times 10^{-7}$} & \multirow{2}{*}{$9.5\times10^{17}$}  & \multirow{2}{*}{7} & $3.75\times 10^{44}$ & \multirow{2}{*}{$4.9\times10^{41}$} & \multirow{2}{*}{$3.75\times10^{47}$}

\\\cline{3-4}\cline{8-8}

 &  & Proton & $9\times 10^{-2}$  &  &   &  &$3.75\times 10^{47}$  &  &     
 
 \\\cline{2-10}
 
 & \multirow{2}{*}{Zone-3} & Electron & $6.1\times 10^{-16}$  & \multirow{2}{*}{$3.98\times 10^{-8}$} & \multirow{2}{*}{$1.2\times10^{22}$}  & \multirow{2}{*}{3} & $7.44\times 10^{40}$ & \multirow{2}{*}{$4.85\times10^{48}$} & \multirow{2}{*}{$4.86\times10^{48}$}

\\\cline{3-4}\cline{8-8}

 &  & Proton & $2.1\times 10^{-11}$  &  &   &  &$2.56\times 10^{45}$  &  &   \\

 \hline

\end{tabular}}
\end{center}
\label{Table-2}
\end{table*}

\par The value of the bulk Lorentz factor is assumed to be 7 in Zone--1 \& Zone--2 and 3 for the extended jet labelled as Zone--3 respectively. The corresponding values of the viewing angle $\theta_{obs}$ are  assumed to be 3\degr in Zone--1 \& Zone--2 and 5.5\degr in Zone--3.
The values of the Doppler factor ($\delta$) are 12.3 \& 5.42 for the respective zones.
\par  As mentioned earlier the maximum energy of protons in the extended jet is $3.98\times 10^{21}$ eV.  \cite{Hillas(1984)}; \cite{Cesarsky(1992)}; \cite{Rachen and Bierman(1993)}; \cite{Henri et al.(1999)} have discussed that the AGN jets can act as potential sites for accelerating protons up to energies of $10^{20}$ eV. According to the findings of \cite{Ebisuzaki and Tajima(2013)} intense electromagnetic fields may give rise to plasma wakefield which can accelerate protons to energies beyond $10^{21}$ eV. The acceleration mechanism in such a case would then be credited to the Lorentz invariant pondermotive force. For our choice of ambient magnetic field \textit{B}=1 mG, the emission region radius of $R=1.2\times 10^{22}$ cm is sufficient for accelerating the protons to $3.98\times 10^{21}$ eV from \autoref{eqn_4}.   We note that  although the maximum energy of protons may be very high from Hillas criterion, when considering  known acceleration mechanisms it is difficult to explain such a high value (\cite{Aharonian et al.(2002)}). 
While selecting the radius ($R=1.2\times 10^{22}$ cm $=3.9$ kpc) of the emission region we have also considered the present estimates of the extended jet dimensions (length--14 kpc, width--4.8 kpc) (\cite{Kaufmann et al.(2013)}). 
\par The three zone modeling of the observed SED of AP Librae has been presented in \autoref{Figure-2}(Bohm diffusion regime), \autoref{Figure-3}(Kolmogorov diffusion regime) \& \autoref{Figure-4}(Kraichnan diffusion regime) along with the relevant parameters used while modeling the data in \autoref{Table-1}. The corresponding power requirement in the three zones is discussed in \autoref{Section_3} along with the other findings.
\par
 The contributions of disc, BLR and torus radiations to the IC emission by the blob and pc-scale jet of AP Librae are considered in earlier papers.
The energy of the BLR photons is blue shifted in the rest frame of the blob. The reprocessed BLR emission has been parametrised in eq(12) of \cite{Hervet et al.(2015)}. The parameter values have been chosen such that (see Table 3)  at high energy external compton emission of blob-BLR origin is significant. Similar discussions are also given in \cite{Zacharias and Wagner(2016)}. Direct emissions from disc, BLR and torus are low as they are not detected although external Compton emission from blob due to the reprocessed emission could be significant (depending on the parameter values chosen, and the distances of the blob and jet from the black hole) after Doppler boosting.
Moreover it is discussed that due to the outward relativistic motion of the blob and its distance from the accretion disc, the external Compton emission of blob electrons due to disc radiation is negligible.

From eqn.(A.6) of \cite{Zacharias and Wagner(2016)} the energy density of reprocessed emission from the dusty torus in the reference frame of the blob is
$u_{DT}=\dfrac{4}{3} \Gamma^2 \dfrac{L_{DT}  \tau_{DT}}{4 \pi  {r_{DT}^{2}} c}$.  Assuming the radius of the region within which the emission is confined to be $r_{DT}=1$pc, luminosity $L_{DT}=10^{42} $ergs/sec, Lorentz factor of blob $\Gamma=7$ and the efficiency of reprocessing $\tau_{DT}=0.001$ the energy density in radiation $u_{DT}=2\times 10^{-8}$ ergs/cm$^3$, which is much lower the synchrotron photon energy density inside the blob $1.28 \times 10^{-5}$  ergs/cm$^3$. By adjusting the parameter values suitably the energy density in reprocessed external radiation could be made even lower. Moreover the pc-jet is assumed to be located beyond the BLR and torus region, which makes external Compton emission insignificant.
 
\section{Results and Discussions}
\label{Section_3}

As shown in \autoref{Figure-2}, \autoref{Figure-3} \& \autoref{Figure-4} the observed quiescent state spectral energy distribution of AP Librae has been fitted by the three zone model adopted in the present work. The optical--UV frequency data in the SED is accounted for partially by the synchrotron emission of electrons from blob and partially by the blackbody radiation from the accretion disk. The thermal emission from the accretion disk is reproduced by assuming a Shakura-Sunyaev type disk (\cite{Shakura and Sunyaev(1973)}) having a luminosity of $L_D = 1.3\times 10^{44}$ ergs/sec. Incidentally \cite{Zacharias and Wagner(2016)} have also used the same estimate for disk luminosity. 

 \begin{equation}
    P_{jet}=\pi R^{2} \Gamma^{2} c (U_{el}+U_{pr}+U_{B})
\label{eqn_11}   
\end{equation}
The jet power requirements from each region is calculated from the expression of $P_{jet}$ presented in \autoref{eqn_11}. 
For the parameters used in our model we find that the required jet power in Zone--1 is $5.7\times10^{47}$ ergs/sec \& $3.75\times 10^{47}$ ergs/sec in Zone--2. The jet power requirement in Zone--3 is $4.86 \times 10^{48}$ ergs/sec. 
 
\par

  Assuming a simple power law proton spectrum of spectral index 2 in the energy range of $10^{15}$--$10^{18}$ eV we have calculated the photon fluxes for the extended jets of 3C 273, PKS 0637-752 and B3 0727+409  to explain the X-ray data with the proton synchrotron model. 
 \begin{figure}[H]
\centering
\includegraphics[width=.5\textwidth]{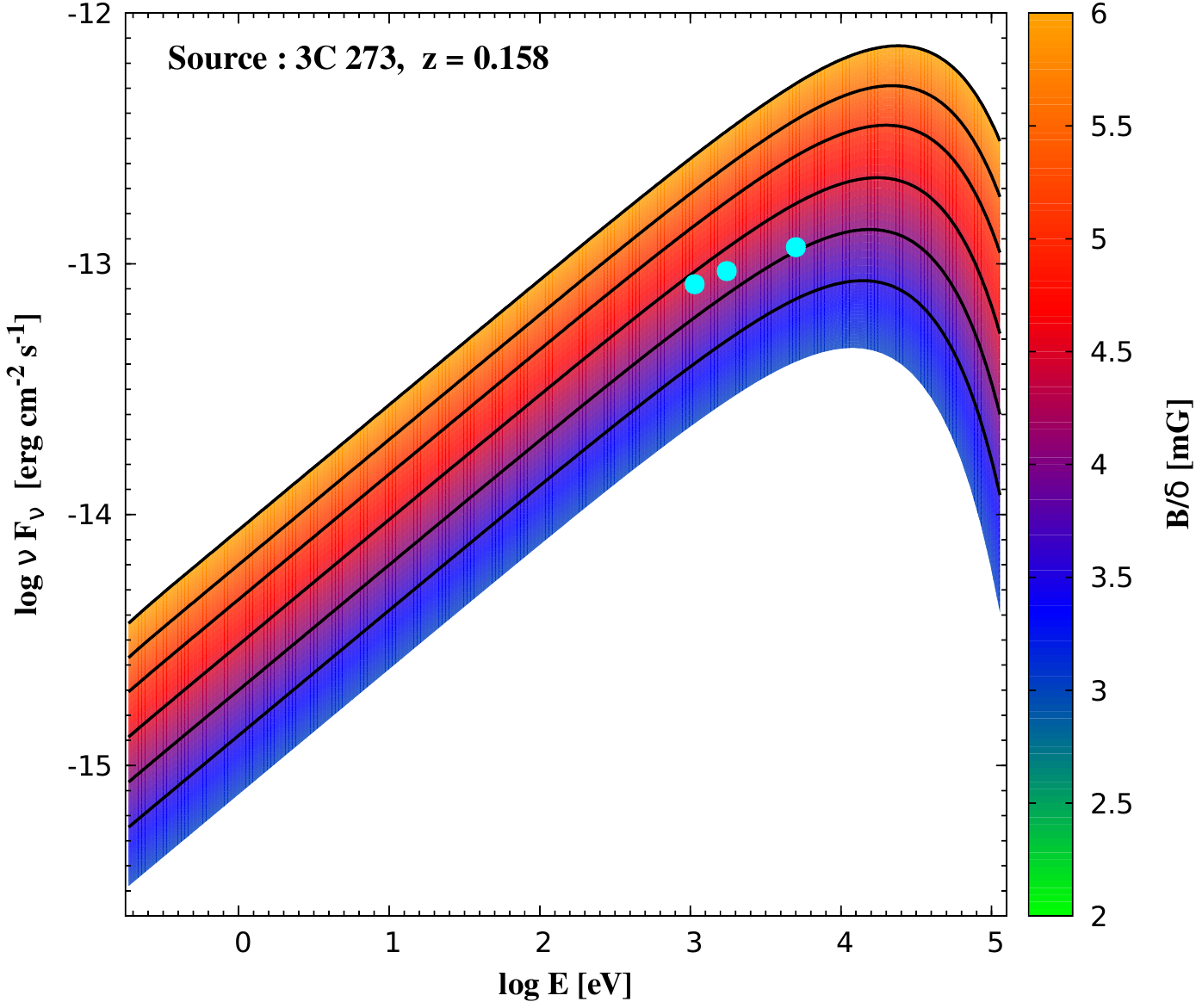}
\caption{\footnotesize{Photon flux from Proton synchrotron emission vs Energy (in logscale) contour plot assuming equipartition of energy in protons and magnetic field. The contour lines are for specific values of the magnetic field $B$ which is equivalent to $\dfrac{B}{\delta}$ (as $\delta$=1 for these plots). The contour lines range from $\dfrac{B}{\delta}$=3.5 mG -- 6 mG in steps of 500 $\mu$G.}}
\label{Figure-5}
\end{figure}
The emission region is assumed to be of radius 1 kpc and its Doppler factor is fixed at 1 for simplicity.
 \autoref{Figure-5}, \autoref{Figure-6} \& \autoref{Figure-7} shows the contour plots for the extended jets of these three sources for magnetic field within 15 mG. We have also assumed equipartition in energy between the magnetic field and the relativistic protons.
\begin{figure}[H]
\centering
\includegraphics[width=.5\textwidth]{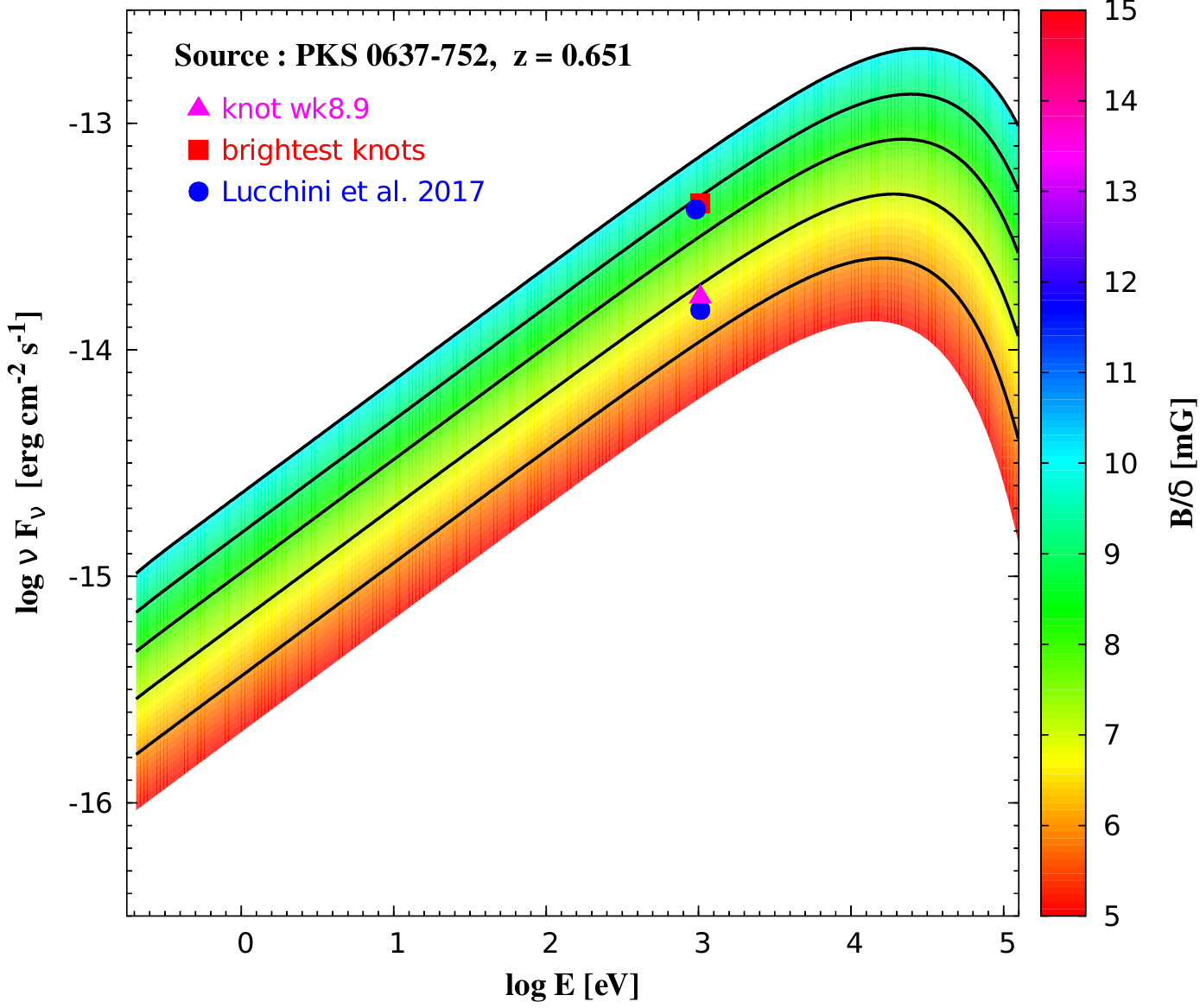}
\caption{\footnotesize{Same as in \autoref{Figure-5} with the source as PKS 0637-752. The contour lines range from $\dfrac{B}{\delta}$=6 mG -- 10 mG in steps of 1 mG. The blue points are taken from \cite{Lucchini et al.(2017)}.}}
\label{Figure-6}
\end{figure}

\begin{figure}[H]
\centering
\includegraphics[width=.5\textwidth]{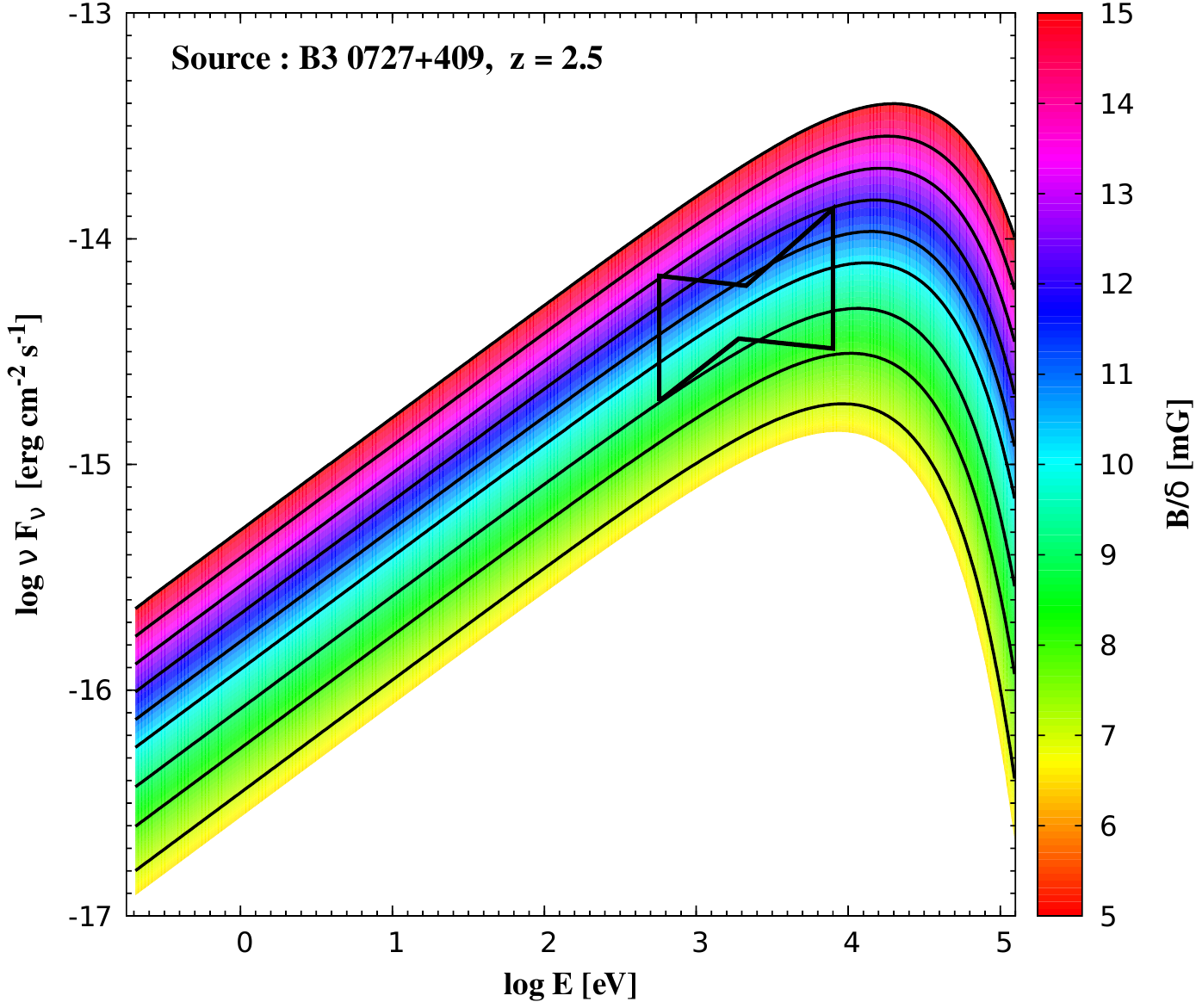}
\caption{\footnotesize{Same as in \autoref{Figure-5} with the source as B3 0727+409. The contour lines range from $\dfrac{B}{\delta}$=7 mG -- 15 mG in steps of 1 mG. }}
\label{Figure-7}
\end{figure}  
      The jet power required to explain the observed X--ray data is $\sim 1.28 \times 10^{48}$ ergs/sec (corresponding to 4.25 mG) for 3C 273,  $\sim 3 \times 10^{48}$ ergs/sec for knot wk8.9 (corresponding to 6.5 mG) \& $\sim 5.75 \times 10^{48}$ ergs/sec for the brightest knot (corresponding to 9 mG) of PKS 0637-752. The Eddington's luminosity of 3C 273  and  PKS 0637-752 could be as high as $10^{48}$ erg/sec (\cite{Paltani and Turler(2005)}, \cite{Kusunose and Takahara(2016)}).  
 Thus the jet powers required to explain the X-ray data  from the extended jets of 3C 273 (for magnetic field 4.25 mG) and PKS 0637-752 (for magnetic fields of 6.5 mG \& 9 mG)  are comparable to their Eddington's luminosities.  Hence it is possible that proton synchrotron emission could be the underlying mechanism of X-ray emission from their extended jets.
 B3 0727+409 is located at a redshift of 2.5 having a black hole mass of $3.3\times 10^8$ solar mass (\cite{Jamrozy et al.(2014)}),
 Eddington's luminosity $4\times 10^{46}$ ergs/sec. In \autoref{Figure-7} the butterfly shows the possible X-ray emission from the extended jet of this source. 
  In this case the jet power required to explain the X-ray emission with proton synchrotron model is $7.1\times10^{48}$ ergs/sec (corresponding 10 mG), which exceeds the Eddington's luminsoity of this source.
  
  \par
  
   We note that due to the VHE $\gamma$-ray emission of  AP Librae it is extremely difficult to model this source assuming equipartition in energy between the magnetic field and protons  after maintaining all the relevant constraints.
  If we lower the magnetic field (at the expense of increasing the energy density of protons) to satisfy equipartition we find that the Hillas criterion can no longer be maintained for the maximum proton energies required by the model. When the radius of the emission region is increased to satisfy the Hillas criterion at equipartition the generated spectra overshoots the observed spectra by a significant amount thus necessitating a departure from equipartition to successfully explain the observed SED. Also a derease in the magnetic field results in a significant increment of the maximum proton energy, which is already extremely high ($3.98\times 10^{21}$ eV) and in tension with the maximum energies predicted by the theoretical models. Thus it is evident that the SED of AP Librae cannot be modelled assuming equipartition on the other hand as the sources 3C 279, PKS 0637-752, B3 0727+409 are not detected in the VHE regime they could be easily modelled with equipartition.  The estimates of jet powers obtained from the study presented in \autoref{Figure-5}, \autoref{Figure-6} \& \autoref{Figure-7} may vary to a certain extent on departure from equipartition but nevertheless they still provide  reasonable benchmarks for the acceptable range of the required jet powers to model the SEDs.

  \par
   The large-scale poloidal and toroidal magnetic fields in MHD jets have been studied earlier (\cite{Vlahakis and Konigl(2004)}, \cite{Komissarov et al.(2007)}, \cite{Romero et al.(2016)}). 
   If magnetic-flux is conserved in conical jets then magnetic energy density decreases as the jet expands. In this case if the magnetic field is purely poloidal then varies as $1/s^2$ and 
   if purely  toroidal  then varies as $1/s$, where $s$ is the distance from the black hole. 
   But in the case of MHD-driven outflow these relations are unlikely to remain valid. In our model also the required magnetic fields in the blob, pc-jet and the extended jet do not obey the $s$ dependence expected from conservation of magnetic flux, indicating the presence of MHD-driven outflow. 

 \par
 Here we note that in earlier papers \cite{Kundu and Gupta(2014)} and  \cite{Bhattacharyya and Gupta(2016)} the luminosity required in the extended jet to explain the observed X-ray data was calculated  by assuming a long jet lifetime of the order of $10^7$--$10^8$ years.
  After dividing the total energy required in the extended jet by the jet lifetime the required luminosity was found to be lower than the Eddington's luminosity.
   \section{Concluding Remarks}
 This paper discusses the possibility of applying the proton synchrotron model to the extended jet of AP Librae to explain the HE and VHE $\gamma$-ray data. Our results indicate that in order to successfully explain the VHE emission from the extended jet of AP Librae via proton synchrotron radiation the following conditions are essential :--\\
    i) the ambient magnetic field in the extended jet must be $\sim 1$ mG.\\
    ii) the existence of extremely high energy protons $>10^{21}$ eV in the extended jet. \\
Although the above conditions are consistent with the Hillas criterion it is evident that the value of the ambient magnetic field required in the proton synchrotron model to explain the extended jet emission is at least an order of magnitude higher than that required in the IC/CMB model. We note that it would be possible to lower the magnetic field if we increase the value of $\Gamma$ from the present value of 3. However, the jet power of the extended jet remains much higher than the Eddington's luminosity of AP Librae. Thus  proton synchrotron model is unlikely to explain the very high energy gamma ray emission from the extended jets of AGN.   
\par

Proton synchrotron model might explain the X-ray emission from the extended jets  of some of the quasars like 3C 273 and PKS 0637-752. In future with many more observations of extended X-ray jets it would be possible to know whether IC/CMB and proton synchrotron model are equally viable models for X-ray emission from extended jets.
 \section{Acknowledgements}
 We are thankful to Michael Zacharias and Stephen Wagner for providing us the data points used in \cite{Zacharias and Wagner(2016)}. 
 We also gratefully acknowledge Ruo-Yu Liu and Felix Aharonian for critical reading of the manuscript and insightful comments. Sincere gratitude is also due to the anonymous referee for a detailed report on the manuscript which helped improve the paper significantly.

\end{document}